\documentclass[twocolumn,aps,prb,showpacs,preprintnumbers,amsmath,amssymb]{revtex4}
\usepackage{graphicx}
\usepackage{bm}
\usepackage{gensymb}
\usepackage{color}
\usepackage{url}

\begin{document}
\title{Tunable site- and orbital-selective Mott transition and quantum confinement effects in La$_{0.5}$Ca$_{0.5}$MnO$_3$ nanoclusters}
\author{A.~Valli$^{1,2}$, H.~Das$^{3,4}$, G.~Sangiovanni$^{5}$, T.~Saha-Dasgupta$^{3}$, and K.~Held$^{1}$}
\affiliation{$^1$ Institute for Solid State Physics, Vienna University of Technology, 1040 Wien, Austria\\ 
$^2$ Democritos National Simulation Center, CNR-IOM 
and Scuola Internazionale Superiore di Studi Avanzati (SISSA), 34136 Trieste, Italy\\
$^3$ S.N. Bose National Centre for Basic Sciences, 700098 Kolkata, India\\
$^4$ School of Applied and Engineering Physics, Cornell University, Ithaca, 14853 New York, USA\\
$^5$ Institute for Theoretical Physics and Astrophysics, University of W\"urzburg, Am Hubland, 97074 W\"urzburg, Germany}



\date{\today}
\begin{abstract}

We present a dynamical mean-field theory (DMFT) study of the charge and orbital correlations 
in finite-size La$_{0.5}$Ca$_{0.5}$MnO$_3$ (LCMO) nanoclusters. 
Upon nanostructuring LCMO to clusters of $3$~nm diameter, the size reduction induces an insulator-to-metal transition 
in the high-temperature paramagnetic phase.  This is ascribed to the reduction in charge disproportionation 
between Mn sites with different nominal valence [Das \emph{et al.}, Phys. Rev. Lett. {\bf 107}, 197202 (2011)]. 
Here we show that upon further reducing the system size to a few-atom nanoclusters, quantum confinement effects come into play. 
These lead to the opposite effect: the nanocluster turns insulating again and the charge disproportionation 
between Mn sites, as well as the orbital polarization, are enhanced. 
Electron doping by means of external gate voltage on few-atom nanoclusters is found to trigger 
a site- and orbital-selective Mott transition. 
Our results suggest that LCMO nanoclusters could be employed for the realization of technological devices, 
exploiting the proximity to the Mott transition and its control by size and gate voltage. 
\end{abstract}
\maketitle

\section{Introduction} Research on manganites dates back to the 1950's, 
when Jonker and van Santen\cite{jonkerPU} reported the existence of ferromagnetic metallic phase 
in mixed crystals of manganese oxides LaMnO$_3$-CaMnO$_3$, LaMnO$_3$-SrMnO$_3$, and LaMnO$_3$-BaMnO$_3$.
However, the interest of a wide portion of the scientific community was only raised in the 1990s, 
due to the experimental observation of a colossal magnetoresistance (CMR) effect. \cite{vonHelmolt93,Jin94}
Indeed, the relative change in resistivity upon the application of an external magnetic field 
was much higher than the one observed in artificial magnetic/non-magnetic multilayer systems: 
up to $60\%$ at room temperature in thin films.\cite{kustersPB,vonHelmolt93} 
Triggering such a CMR, however, requires cooling below the Curie temperature $T_C$ 
and the application of relatively strong magnetic fields, preventing the technical application of the CMR effect to this day. 
Alternative routes to achieve CMR have also been followed 
in mixed valence manganites such as the half-doped La$_{0.5}$Ca$_{0.5}$MnO$_3$ (LCMO), \cite{Schiffer95}
which is insulating below $155$~K and displays antiferromagnetic and charge order, \cite{wollanPR100} 
often also accompanied by orbital order. 
Indeed, the antiferromagnetic insulating state is prone to instabilities. 
The transition toward a ferromagnetic metallic state can be triggered upon applying 
a magnetic field,\cite{kuwaharaSci270} doping, biaxial strain, pressure,\cite{kozlenkoJPCM16} or an electric field.\cite{asamitsuNat388} 
The experiments \cite{sarkarJAPL92,sarkarPRB77,zhouJAP107,chowdhuryNR15,iniamaJAP116} 
suggest that the destabilization of the charge-orbital order can also be obtained upon size reduction. In fact, it was observed also in Nd$_{1-x}$Ca$_{x}$MnO$_{3}$ \cite{raoPRB74,bhagyashree1412.2450} 
and Sm$_{1-x}$Ca$_{x}$MnO$_{3}$ \cite{goveasJAP117} in both at half-doping and in asymmetrically doped samples, 
as well as in Pr$_{0.5}$Ca$_{0.5}$MnO$_{3}$ \cite{raoAPL87,sarkarJAP101,zhangPRB80} compounds. Nonetheless,
there remains some controversy \cite{jirakPRB81} 
arising from the experimental difficulty to disentangle the effects of size reduction 
from other effects, such as oxygen non-stoichiometry, disorder, strain etc.

On the theoretical side,  the effects of size reduction has been studied by 
density functional theory (DFT)+U  \cite{dasPRL107,dasPRL107SM,tangJMMM333} and DFT+dynamical mean-field theory (DMFT). \cite{dasPRL107}
The theoretical analysis is in remarkable agreement with the experiments and shows a correlation-driven destabilization
of the charge-orbital order in bulk LCMO upon size reduction. 
This bears the prospects that, for the optimized size, a much smaller
magnetic field is  sufficient to trigger a CMR, as 
LCMO nanoclusters can be tuned to  the verge of a metal-insulator transition.
Hitherto, the DFT+DMFT calculations on this topic, e.g., those reported in Ref.\ \onlinecite{dasPRL107},
were performed for bulk model systems with \emph{ab-initio} parameters. 
That is, following the DFT calculations for $3$~nm clusters, model bulk systems were constructed 
having the same unit cell volume as well as octahedral distortion as in the core of $3$~nm cluster. 
This way, it was possible to take into account the interplay between strong electronic correlations within DMFT 
and the structural distortions induced by size reduction, obtained \emph{ab-initio} through atomic relaxation within DFT. 
In this paper, we take a significant step forward, in terms of carrying out nanoscopic DMFT calculations 
for few-atom clusters with a DFT-derived tight-binding Hamiltonian. This gives us the opportunity to consider 
the effect of size reduction from bulk to intermediate-sized clusters to few-atom clusters. 
Our calculations show an interesting evolution from the high-temperature paramagnetic insulating (PI) state in the bulk 
to a paramagnetic metallic (PM) state in intermediate-sized clusters to re-entrant insulating solution for few-atom clusters. 
We also investigate the effects of applied hydrostatic pressure in the bulk, 
which turn out to be different than the effects of size reduction. 
Considering few-atom clusters, we also show that electron doping, 
through the application of external gate voltage, drives an unexpected site- and orbital-selective Mott transition.  

The paper is organized as following. 
In Section \ref{sec:method} we discuss the model employed for the description of mixed-valence manganites, 
and we describe the strategy we followed to include structural, finite-size, and many-body effects 
in the framework of a combined DFT+DMFT approach. 
In Sec. \ref{sec:results} we present the DMFT results obtained for LCMO nanoclusters of different size. 
In particular, in Sec. \ref{sec:spp} we focus on the effects of quantum confinement 
on the spectral properties, while in Sec. \ref{sec:gate} we explore the effect of electrostatic doping 
by applying an external gate voltage to the few-atom clusters. 
Finally, in Sec. \ref{sec:conclusions} we present our conclusions.

\section{Method: DFT+DMFT approach for La$_{0.5}$Ca$_{0.5}$MnO$_3$ nanoclusters}
\label{sec:method}
 
\subsection{Bulk crystal and electronic structure}
\label{sec:structure}

Manganites, R$_{1-x}$A$_x$MnO$_3$ with R being a trivalent rare-earth-metal element and A  a divalent dopant,
have a perovskite lattice structure, with the rare-earth atoms at corner positions, 
the Mn atoms at body center positions and oxygen atoms at the face centered positions.
Depending on the sizes of R and A, given by the so-called tolerance factor,
the MnO$_6$ octahedra can tilt and rotate reducing the symmetry of the perovskite lattice from cubic to orthorhombic.
With nominal oxygen valency O$^{2-}$, in half-doped compounds (i.e., $x=0.5$) 
the manganite atoms are in a mixed valent Mn$^{3.5+}$ state. 
This can lead to a charge disproportionation between the Mn sites in bulk half-doped manganites; 
in the extreme case one has a 50\%  of Mn$^{4+}$ sites with a $3d^3$ configuration, 
and the other 50\% of Mn$^{3+}$ sites with a $3d^4$ configuration. 
The charge-ordered state is associated with a real space ordering of Mn$^{3+}$/Mn$^{4+}$ species in a 1:1 pattern. 
The octahedral crystal field surrounding the Mn ions splits the $3d$ orbitals 
into three low energy $t_{2g}$ orbitals ($d_{xy}$, $d_{xz}$, and $d_{yz}$) 
and two higher energy $e_g$ orbitals ($d_{3z^2-r^2}$ and $d_{x^2-y^2}$). 
Due to the strong Hund's exchange coupling, the three Mn electrons of Mn$^{4+}$ 
occupy the $t_{2g}$ orbital, in a high-spin $S_{t2g}\!=\!3/2$ configuration. 
The consequent energy gain associated to this state, makes it energetically favorable 
for the extra electron of Mn$^{3+}$ to occupy the doubly degenerate $e_g$ states. 
Hence, the charge ordering accommodates a static Jahn-Teller distortion at the Mn$^{3+}$ sites, 
removing the degeneracy and lowering the symmetry of the system to monoclinic. 

The Mn $e_g$ orbitals are delocalized due to a strong hybridization with the O$2p$ states. 
On the contrary, Mn $t_{2g}$ orbitals do not hybridize strongly with the O and are localized. 
The magnetism is thus governed by the double exchange mechanism.\cite{zenerPR82} 

\subsection{DFT calculations and downfolding procedure}
\label{sec:DFT}

The first step of our study is a DFT calculation of both LCMO bulk and nanoclusters. 
To this end, we use projected augmented wave (PAW) pseudopotentials 
with an energy cutoff of $450$~eV and performed calculations 
within a spin-polarized generalized gradient approximation (GGA) \cite{perdewPRL77} as implemented 
in the Vienna \emph{ab-initio} Simulation Package (VASP). \cite{kressePRB47,kressePRB49,kresseCMS6,kressePRB54} 
The forces on the atoms are converged to less than $0.01$ eV/\AA. 
All DFT calculations for bulk LCMO was performed with a $4 \times 2 \times 4$ $k$-mesh. 
On the other hand, for the nanoclusters we use only the $\Gamma$ point for the $k$-space integration. 
The structural optimization of the bulk LCMO is performed considering 
both antiferromagnetic (AFM) CE and ferromagnetic (FM) configurations. 
We find  the AFM CE phase to be lower in energy compared to the FM phase by $45$ meV/f.u. 
The stoichiometry of the system is maintained in the construction of the nano-clusters, 
as discussed in Refs. \onlinecite{dasPRL107} and \onlinecite{dasPRL107SM}. 

A necessary step for the subsequent DMFT calculation is the extraction 
of the relevant tight-binding parameters, calculated {\em ab initio} within DFT. 
To achieve this, we employ the downfolding method 
as implemented in the N-th order muffin tin orbital (NMTO) basis \cite{andersenPRB12,andersenPRB62} 
with potential parameters borrowed from self-consistent 
linearized muffin-tin orbital (LMTO) \cite{andersenPRB12} calculations. 
Through the NMTO-downfolding procedure, a low-energy Hamiltonian $H(k)$ 
involving only Mn $e_g$ Wannier orbitals\cite{wannierPR52} is constructed in $k$-space 
by integrating out all other degrees of freedom. 
The Fourier transformation of $H(k)$ provides the tight-binding parameters. 
In the following we refer to this structure as $S_{\rm bulk}$. 
We use $S_{\rm bulk}$ to describe bulk manganites within standard DFT+DMFT calculations. 
In order to construct a low-energy Hamiltonian for the nanoclusters, as explained in Ref.\ \onlinecite{dasPRL107}, 
a cluster of nearly spherical shape having $3$~nm diameter is cut out from a large supercell 
of the bulk crystal structure in monoclinic P2$_{1}$/m symmetry, which is then subject to a full structural optimization. 
In the following we refer to this structure as $S_{\rm nano}$. 
The NMTO-downfolding calculation is then carried out on the self-consistent LMTO calculation 
for a model bulk structure, referred to as $S_{\rm model}$, which is constructed considering the MnO$_6$ octahedra 
selected from the core region of $S_{\rm nano}$, 
and applying various symmetry operations, as explained in detail in Ref.\ \onlinecite{dasPRL107SM}. 
The DF-derived tight-binding Hamiltonians contain the information of the structural as well as electronic 
changes at the level of one-electron theory that happen upon size reduction. 

In order to take into account many-body effects within DMFT, we build a low-energy model 
(discussed in detail in Sec.\ \ref{sec:Hlowe}) with the DFT-derived hopping parameters as an input. 
However, there is fundamental difference between the calculations performed in this work 
and that of Ref.\ \onlinecite{dasPRL107}, as explained is the following: 
In Ref.\ \onlinecite{dasPRL107}, we performed DMFT calculations 
on the constructed model \emph{bulk} structure ($S_{\rm model}$). 
Instead, in the present work, we use the same parameters of $S_{\rm model}$ to construct few/atom nanoclusters of different size, 
which we solve within the nano-DMFT scheme discussed in Sec.\ \ref{sec:DMFT}.


\subsection{Low-energy effective  $e_g \!+\!S_{t_{2g}}$ model}
\label{sec:Hlowe}

The low-energy Hamiltonian describing the manganites is given as below,\cite{helddv,ahnPRB61,yamasakiPRL96}
\begin{equation} \label{eq:HamGGA+DMFT}
 \begin{split}
  {\cal H} & \!=\! \sum_{ij mm'}\sum_{\sigma\sigma'} h_{ij,mm'} c^{\dagger}_{im\sigma} c^{\phantom{\dagger}}_{jm'\sigma'} 
                  \!-\! {\cal J}S \sum_{im}(n_{im\uparrow} \!-\! n_{im\downarrow}) \\
               & \!+\!  U \sum_{im} n_{i m \uparrow} n_{i m \downarrow} 
                  \!+\! \sum_{imm'}\sum_{\sigma \sigma'} (U'\!-\!J\delta_{\sigma\sigma'}) n_{im\sigma}n_{im'\sigma'}.
 \end{split}
\end{equation}
Here, $h_{ij,mm'}$ denotes the generic matrix element of the one-particle DFT Hamiltonian 
in the basis of the the downfolded NMTO Wannier orbitals $m$, $m'$ at site $i$, $j$; 
the on-site Coulomb interaction between the $e_g$ electrons is parametrized 
in terms of an intra-orbital repulsion $U\!=\!5$~eV, 
a Hund's exchange $J\!=\!0.75$~eV, and an inter-orbital interaction $U'\!=\!U\!-\!2J$. \cite{note_intpar, heldAP56} 
The values of the interaction parameters, taken from the literature, \cite{parkPRL76} 
represent realistic estimates for manganites. 
Furthermore, the $e_g$ orbitals are coupled to a (classical) disordered spin $S$ 
by a Hund's exchange ${\cal J}$. This term represents the interaction 
between the itinerant $e_g$ electrons with the localized electrons in the half-filled $t_{2g}$ manifold. 
In the notation adopted here, $S \!=\! \pm 1$ is the classical spin corresponding to 
the high-spin states of $S_{t_{2g}}\!=\!\pm3/2$ while its modulus  
is included in the value ${\cal J}\!=\!1.35$~eV.\cite{yamasakiPRL96} 
Hamiltonian (\ref{eq:HamGGA+DMFT}) represents a standard low-energy model for the description 
of electronic correlations in manganite compounds \cite{helddv,ahnPRB61,yamasakiPRL96} 
Recently, the classical spin description of the $t_{2g}$ manifold has been thoroughly revisited. \cite{fleschPRB87}  
Thanks to the significant technical improvements in the field of impurity solvers, this allows the direct treatment 
of a five-orbital model for the whole Mn$3d$ multiplet, including also {spin-flip} and {pair-hopping} terms 
beyond the density-density Coulomb interaction. \cite{gullRMP83,parraghPRB86} 
Remarkably, in the case of pure LaMnO$_3$, a detailed analysis showed an excellent agreement 
between a classical spin and a full quantum many-body treatment of the $t_{2g}$ orbitals, 
e.g., for the $e_g$ spectral functions. \cite{fleschPRB87} 
Moreover, the estimate obtained for the spin-spin correlations functions $\langle S_z^{e_g} S_z^{t_{2g}} \rangle \!\approx\!0.74$  
was found to be consistent with the picture of aligned $e_g$ and $S\!=\!3/2$ $t_{2g}$ spins. \cite{fleschPRB87} 
In light of these considerations, the restriction to the low-energy model described by Hamiltonian (\ref{eq:HamGGA+DMFT}) 
represents a realistic and physically sensible choice to study correlation effects 
in Mn compounds with a half-filled $t_{2g}$ manifold and partially filled $e_{g}$ orbitals.

\subsection{Dynamical mean-field theory with inequivalent Mn atoms in the unit cell}
\label{sec:DMFT}

In the following we discuss the technical details for the solution of the many-body Hamiltonian (\ref{eq:HamGGA+DMFT}) 
in the framework of DMFT \cite{georgesRMP68,heldAP56} for inhomogeneous systems. \cite{potthoffPRB59,snoekNJP10,valliPRL104,valliPRB86,titvinidzePRB86,valliPRB91}
%
That is, we  solve an auxiliary Anderson impurity problem for each inequivalent Mn site in the unit cell. 
Moreover,  we perform an average over the disordered classical $t_{2g}$ spin $S$. \cite{note_CPA}
This procedure yields a local $2\times2$ self-energy in the $e_g$ manifold of each of the Mn atoms in the unit cell, 
while neglecting non-local self-energy elements between different Mn atoms. 
Let us start by discussing the bulk DFT+DMFT calculations, which have been performed 
on $S_{\rm bulk}$ and $S_{\rm model}$ structures to describe LCMO bulk and nanoscopic ($3$~nm) clusters. 
The bulk monoclinic unit cell of LCMO contains eight Mn atoms
but only three kinds of Mn atoms are locally inequivalent, labeled as Mn$1(1)$, Mn$1(2)$, and Mn$2$. 
The four Mn$1$ atoms have a nominal valence $3+$ and occupy the bridge sites of the zig-zag ferromagnetic chain 
in the CE-type magnetic order that sets in below the N\'{e}el temperature in bulk LCMO. 
Those Mn$1$ are further divided into two Mn$1(1)$ and two Mn$1(2)$ sites, by symmetry. 
The four Mn$2$ atoms have a nominal valence $4+$ and occupy the corner sites of the zig-zag ferromagnetic chain. 
Exploiting these symmetries, the overall computational effort for the bulk amounts to the solution 
of three auxiliary impurity problems within a DMFT self-consistent scheme, 
with the corresponding subtraction of three inequivalent double counting terms in $h_{ij,mm'}$: 
\begin{equation}
 \Delta^{DC}_{im}\!=\!\tilde{U} \Big(n^{\text{DFT}}_{im} - \frac{1}{2} \Big),
\end{equation}
where $\tilde{U}\!=\!U\!-\!\frac{5}{3}J$ denotes an averaged interaction and $n^{\text{DFT}}_{im}$ 
are the NMTO orbital occupancies for each of the inequivalent Mn in the unit cell. \cite{anisimovPRB44,czyzykPRB49,heldAP56} 
In the numerical calculations, we employ a Hirsch-Fye Quantum Monte Carlo \cite{Hirsch86a} 
impurity solver, with a Trotter discretization $\Delta\tau^2\!\approx\!0.027$ 
and inverse temperature $\beta\!=\!20$ eV$^{-1}$ in the paramagnetic phase. 

\begin{figure}
\includegraphics[width=0.46\textwidth, angle=0]{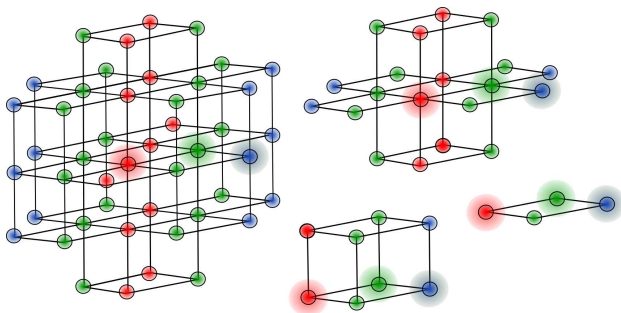}
\caption{(Color online) Schematic representation of the Mn sublattice of  the few atom sized small
clusters considered in the present DMFT work, having $N\!=\!46$, $20$, $8$, and $4$ Mn sites.
The clusters are built by chopping off atoms from the supercell  
described by the tight-binding Hamiltonian of $S_{\rm nano}$.  
The red (gray), blue (dark gray) and green (light gray) atoms correspond to 
Mn$1(1)$, Mn$1(2)$, and Mn$2$, respectively, according to their 
classification in the bulk.  The cluster boundaries correspond to dangling Mn-Mn bonds. 
La, Ca, and O atoms (not shown) are effectively taken into account in the \emph{ab-initio} parameters of $S_{\rm nano}$
through the downfolding procedure.} 
\label{fig:SnanoN}
\end{figure} 

\begin{figure*}
\includegraphics[width=0.95\textwidth, angle=0]{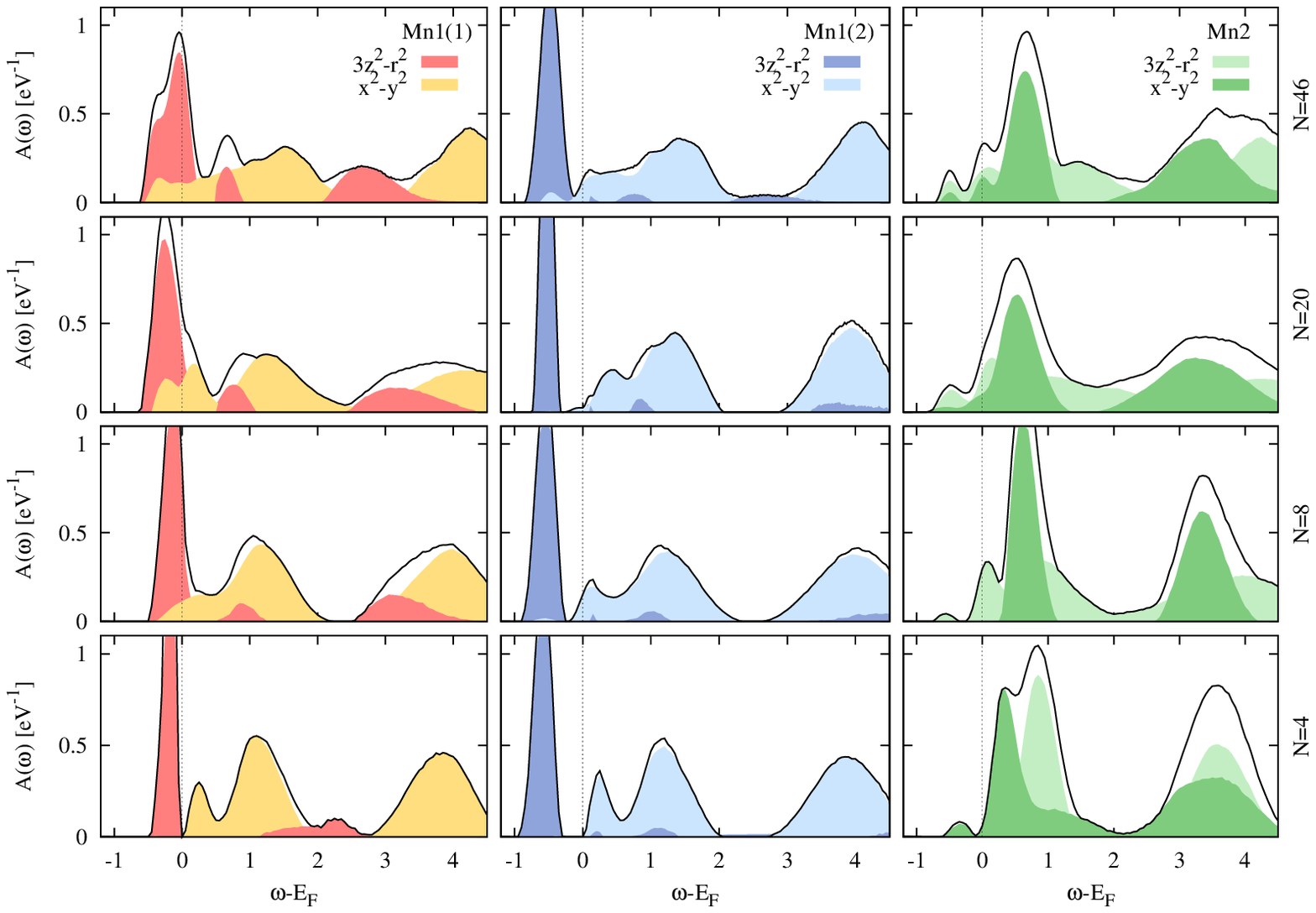} 
\caption{(Color online) Evolution of the spectral properties of the
nanoclusters, shown in Fig. 1, described by the tight-binding parameters of 
$S_{\rm nano}$. Each row shows the spectral function $A(\omega)$ for a cluster
having $N$ Mn atoms. The red (gray), blue (dark gray) and green (light gray) lines 
correspond to representative Mn$1(1)$, Mn$1(2)$, and Mn$2$ atoms, respectively,
indicated by shaded spheres in the nanoclusters shown in Fig.\ \ref{fig:SnanoN}.
The dark and light filled curves in each panel denote the contributions
from the $3z^2-r^2$ and $x^2-y^2$ orbitals,
while the black solid line denotes the on-site spectral density of the
$e_g$ manifold. A metal-to-insulator transition (MIT) upon decreasing cluster size, due to
quantum confinement effect, is evidenced accompanied by an enhancement of the charge and orbital order.}
\label{fig:nano-DMFTspectra_HAMR46}
\end{figure*}

For the nano-DMFT calculations, we consider the LCMO finite-size nanoclusters 
shown in Fig.\ \ref{fig:SnanoN} and described by the structure $S_{\rm nano}$. 
The symmetry of the nanoclusters is much lower than in the bulk due to finite-size effects. 
This leads to many more inequivalent Mn. For example, the $N\!=\!46$ Mn atom cluster contains $N_{\rm{ineq}}\!=\!23$ 
inequivalent Mn sites, as it possesses only the inversion symmetry with respect to the center of the cluster. 
In this case, we need to solve for $N_{\rm{ineq}}$ impurity problems, 
yielding $N_{\rm{ineq}}$ $2\!\times\!2$ local self-energy matrices $\Sigma^{ii}_{mm'}$. 
From these, new $(N\!\times\!2)\!\times\!(N\!\times\!2)$ cluster Green's functions are calculated 
by solving the Dyson equation, enforcing the self-consistency at the level of the whole nanocluster. 
This scheme has been employed for all the few-atom Mn clusters, shown in Fig.\ \ref{fig:SnanoN},
taking into account the specific symmetries of each structure. 
We stress once more that this nano-DMFT approach is different in spirit from that adopted in Ref.\ \onlinecite{dasPRL107}, 
in which bulk-DMFT calculations were performed, and the effect of size reduction was considered 
via the change in tight-binding parameters of $S_{\rm model}$ with respect to $S_{\rm bulk}$. 
Instead, within nano-DMFT calculations we explicitly take into account the boundary effects 
of the finite-size nanoclusters constructed with the tight-binding parameters extracted from $S_{\rm nano}$.

\section{Results for finite size LCMO nanoclusters} 
\label{sec:results}

\subsection{Spectral properties, charge and orbital order}
\label{sec:spp}

In the following we discuss the spectral properties of the nano clusters shown in Fig.\ 1. 
The DMFT spectral functions are shown in Figs. \ref{fig:nano-DMFTspectra_HAMR46}, for
three representative Mn atoms belonging to Mn$1(1)$, Mn$1(2)$ and Mn$2$ kinds, 
following the bulk classification (even though for the nanoclusters the atoms of e.g.  Mn$1(1)$ type are of course not equivalent any longer as in $S_{\rm model}$). 
At the outset, we notice that the size reduction has profound consequences 
on the electronic structure of the LCMO nanoclusters, 
in the sense that the metallic character decreases gradually in moving from $N\!=\!46$ to $4$. 
In the extreme case of $N\!=\!4$, the system is insulating 
even at the high temperature  considered here ($\beta\!=\!20$ eV$^{-1}$), 
reflecting the strong quantum confinement effects induced upon size reduction. 
The observed metal-to-insulator transition is accompanied by a strong enhancement 
of the charge disproportionation between Mn$^{4+}$ and Mn$^{3+}$. 
We further find that size effects are accompanied by an overall enhancement of the orbital polarization. 
These effects are strong for Mn$1(1)$ and Mn$2$ atoms
which are located at the core of the clusters. 
Instead, all Mn$1(2)$ sites are located at the surface of the clusters (with maximum number of dangling bonds) 
and have a nearly insulating and almost fully orbitally-polarized spectral function irrespective of cluster size. \\
In the bulk LMCO, the orbital polarization between two $e_g$ orbitals, $3z^{2}-r^{2}$ and $x^{2}-y^{2}$
is complete for Mn$1$ atoms with nominal $3+$ valence, and zero for Mn$2$ atoms with nominal $4+$ valence.
Moving to cluster of $3$~nm diameter it was shown\cite{dasPRL107} that both charge disproportionation between
Mn$1$ and Mn$2$ atoms, as well as the orbital polarization at the Mn$1$ atoms 
decreases considerably compared to the bulk, leading to metallicity. 
Indeed, we find that upon further reduction of the size to few-atom nanoclusters, 
quantum confinement effect comes into play, making the charge disproportionation and
orbital polarization increase again, especially for Mn$1(1)$ atoms. 
Both charge disproportionation and orbital polarization show an increasing trend 
upon decreasing the cluster size, from $N\!=\!46$ to $20$ to $8$ 
until it becomes almost complete for the $N\!=\!4$ nanocluster. 

In order to study quantitatively the effects of size reduction, we consider the 
cluster averaged charge order and the orbital polarization. 
To this end, we define the occupation $n_{i}$ and the polarization $p_i$ of Mn site $i$ as 
\begin{equation}
\begin{split}
 n_i  & =  \frac{1}{2}\sum_{\{S\}} \sum_{m\sigma} n^{S}_{im\sigma}, \\
 p_i  & =  \frac{1}{2}\sum_{\{S\}} \sum_{m\sigma} n^{S}_{mi\sigma} (-1)^m ,  
\end{split}
\end{equation}
where the average over $\{S\}$ takes into account the two possible configurations of the classical $t_{2g}$ spin $S$ 
(as disused in Sec.  \ref{sec:DMFT}). 
Hence, using the definitions above, we compute the following quantities   
\begin{equation}  \label{Eq:OPCO}
\begin{split}
\langle \Delta n \rangle & = \frac{1}{N_{\alpha}} \sum_{i \in \alpha} n_i - \frac{1}{N_{\beta}} \sum_{i \in \beta} n_i , \\
 \langle \Delta p_{\alpha} \rangle & = \frac{1}{N_{\alpha}} \sum_{i \in \alpha} p_i ,
\end{split}
\end{equation} 
where $\alpha \neq \beta$ denotes Mn$1$ and Mn$2$ sites, respectively, while $N_{\alpha}$, $N_{\beta}$ denote the number of Mn sites 
of the corresponding kind in the nanocluster. 
In particular, $\langle \Delta n \rangle$ is the charge disproportionation averaged over the nanocluster, 
while $\langle \Delta p_{\alpha} \rangle$ is the orbital polarization averaged over all Mn site in the nanocluster 
belonging to the same kind $\alpha$. 
Note that in Eq.\ (\ref{Eq:OPCO}), the occupation $n_i$ and the polarization $p_i$ 
for the different Mn sites $i$ (even those belonging to the same Mn kind $\alpha$) 
are in general inequivalent because of the symmetry of the finite nanoclusters constructed from $S_{\rm nano}$, 
which are different from the bulk models described by $S_{\rm bulk}$ or $S_{\rm model}$. 

\begin{figure}[tt]
\includegraphics[width=0.50\textwidth, angle=0]{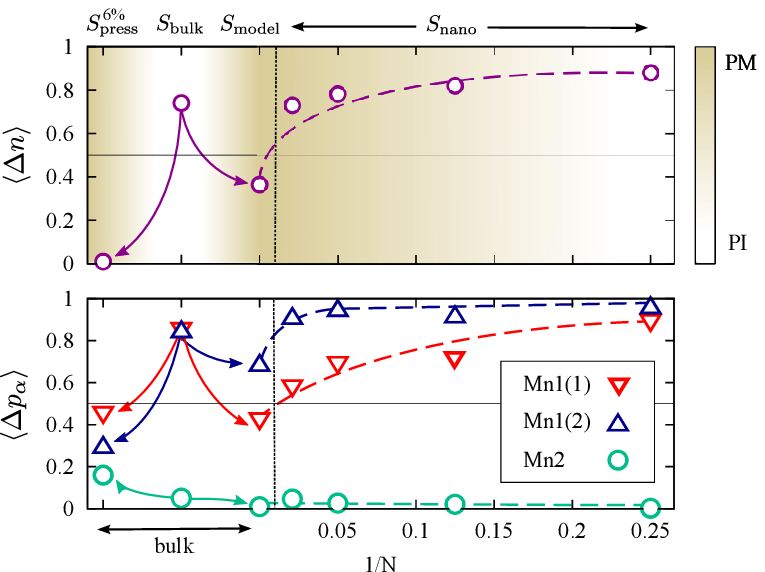}
\caption{ (Color online) Cluster averaged charge order $\langle \Delta n \rangle$ (upper panel) 
and orbital polarization $\langle \Delta p_{\alpha} \rangle$ (lower panel) 
of LCMO bulk and clusters. 
On the right-hand side of each panes we show the nano-DMFT results obtained for the finite-size clusters 
built with the parameters of $S_{\rm nano}$, as a function of the (inverse) number of the Mn atoms in the cluster. 
On the left-hand side of each panel we also show the DFT+DMFT results obtained 
within the calculations of Ref.\ \onlinecite{dasPRL107} for the bulk model $S_{\rm model}$ 
(applicable for $3$~nm diameter cluster containing about $200$ Mn atoms) and on bulk LCMO $S_{\rm bulk}$. 
Those are compared to additional DFT+DMFT result for $S_{\rm press}$, corresponding to the bulk 
under the application of hydrostatic pressure, resulting into 6$\%$ volume reduction. 
The light brown (light tray) shade in the upper panel represents the transitions between PI and PM phases. 
Arrows and dashed lines are a guide to the eye. }
\label{fig:CO}
\end{figure} 

\begin{table}[t]
\caption{Orbital  DFT+DMFT occupancies for the three inequivalent Mn atoms in the unit cell 
of $S_{\rm{model}}$ and $S_{\rm{press}}$. In brackets we give the corresponding occupancies 
for the one-particle low-energy DFT Hamiltonian without the effect of DMFT correlations, 
which strongly enhance the charge-orbital order. Both the orbital polarization in Mn$1$ atoms, 
as well as the charge disproportionation between Mn$1$ and Mn$2$ atoms are more pronounced 
for the nano model $S_{\rm{model}}$  than for the pressurized bulk system $S_{\rm{press}}$. }
\label{tab:nbulk}
\vspace{0.1cm}
\begin{tabular}{l|c|c}
\hline \hline
 &  Pressure 6\% & Nano model \\
\hline \hline
\enskip \enskip  &  $3z^{2}-r^{2}$  \enskip \enskip  $x^{2}-y^{2}$  \enskip \enskip &   $3z^{2}-r^{2}$  \enskip \enskip $x^{2}-y^{2}$ \enskip \enskip \\
Mn1(1) \enskip \enskip & \enskip 0.52 (0.39) \enskip 0.06 (0.11) \enskip \enskip & \enskip 0.52 (0.31) \enskip 0.09 (0.20)  \\ 
Mn1(2) \enskip \enskip & \enskip 0.36 (0.34) \enskip 0.07 (0.13) \enskip \enskip & \enskip 0.72 (0.38) \enskip 0.04 (0.19)  \\
Mn2     \enskip \enskip & \enskip 0.33 (0.31) \enskip 0.17 (0.21) \enskip \enskip & \enskip 0.16 (0.21) \enskip 0.16 (0.25)  \\ 
\hline \hline
\end{tabular}
\end{table}

The cluster-averaged $\langle \Delta n \rangle$ and $\langle \Delta p_{\alpha} \rangle$ 
are shown in the right-hand side of each panels in Fig. \ref{fig:CO} 
as a function of the (inverse) number of the Mn atoms in the cluster. 
For comparison, on the left-hand side of each panel the results 
obtained in Ref.\ \onlinecite{dasPRL107} within DFT+DMFT calculation performed on the bulk nanomodel $S_{model}$  are also shown
(applicable to $3$~nm diameter cluster containing about $200$~Mn atoms). 
As a general trend, the charge order and orbital polarization are found to be the largest for the smallest cluster 
and decrease upon increasing the cluster size. 
In particular, $\Delta n$  shows a smooth reduction with increasing system size 
from a value close to $\Delta n \!=\!1$ for $N\!=\!4$, corresponding to the limit 
in which the Mn$2$ sites are completely empty, to $\Delta n \approx 0.70$ for $N\!=\!46$. 
Those values can be compared to $\Delta n \approx 0.35$ found for the $S_{\rm{nano}}$ model. 
A similar behavior is found for the averaged orbital polarization, 
although the behavior is somewhat dependent on the system shape and symmetry, 
especially for the smaller nanoclusters considered here. 
However, in general the  orbital polarization of the Mn$1$ atoms tends to decrease, by increasing cluster size, 
while that for the (almost empty) Mn$2$ atoms is always negligible. 
A more careful analysis (not shown) reveals that the averaged $\Delta p$ is systematically larger 
than the one of the core Mn atoms (especially for larger clusters), 
as it also includes the contribution of Mn atoms on the surface of the cluster, 
which are characterized by sharper spectral structures and a stronger orbital polarization. \\
If we also consider the results obtained from the DFT+DMFT calculations of Ref.\ \onlinecite{dasPRL107}, 
we find that both the charge order $\langle \Delta n \rangle$ and the orbital polarization $\langle \Delta p_{\alpha} \rangle$ 
are strongly enhanced for $S_{\rm bulk}$ with respect to the $3$~nm cluster size described by $S_{\rm model}$, 
which drives a metal-to-insulator transition. \cite{dasPRL107} \\
The above observations are depicted by the following scenario: 
Starting from the smallest size cluster and upon progressively increasing the system size, 
one encounters an insulator-to-metal transition in few-atoms nanoclusters, 
which is driven by a weakening of quantum confinement effects and charge-orbital correlations. 
With increasing system size, the nanocluster  smoothly evolves towards the results obtained with the bulk nanomodel, 
which is metallic due to the weak structural distortions included in the parameters of $S_{\rm model}$. 
On the other hand, bulk LCMO $S_{\rm bulk}$ is strongly distorted. 
The structural distortions lead to the enhancement of charge-orbital order and drive the system 
across a second metal-to-insulator transition between $S_{\rm model}$ ($3$~nm) and $S_{\rm bulk}$ (bulk). \cite{dasPRL107}

It is interesting to compare the results obtained for $S_{\rm model}$ and for the bulk 
under applied hydrostatic pressure $S_{\rm press}$ corresponding to the same volume reduction of the unit cell, 
i.e., 6\% for $3$~nm nanocluster described by $S_{\rm model}$. 
We find that the effects of hydrostatic pressure and the size reduction are rather different. 
The results for $S_{\rm press}$ indicate that charge order and orbital polarization 
are strongly reduced with respect to the case of $S_{model}$, as shown in detail in Table \ref{tab:nbulk}.   
In particular, we find that for $S_{\rm press}$ the charge disproportionation between Mn$^{3+}$ and Mn$^{4+}$ 
is negligible, and it is accompanied by an overall reduction of the orbital polarization. 
This makes the system under hydrostatic pressure to be far more metallic compared to a $3$~nm cluster.

\begin{figure*} 
 \includegraphics[angle=0, width=0.90\textwidth]{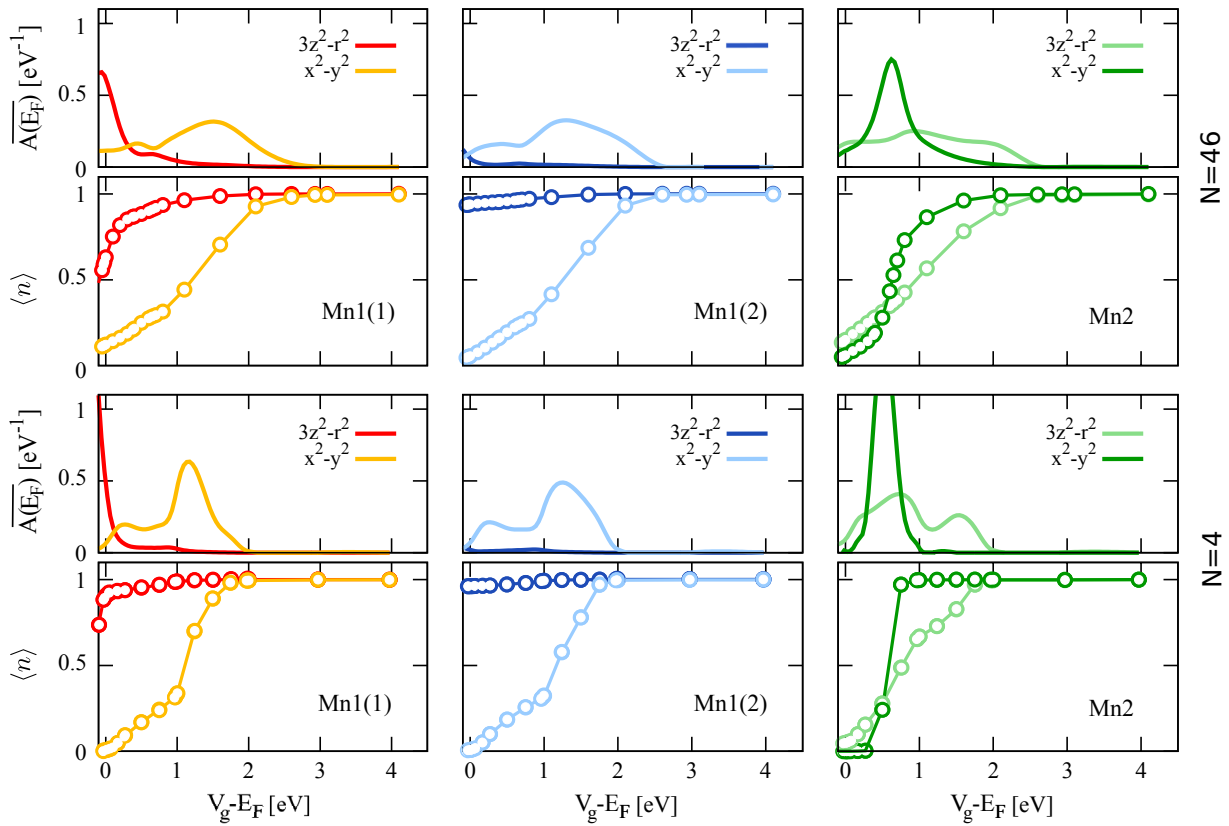}
 \caption{(Color online) Evolution of the spectral weight at the Fermi energy $\overline{A(E_F)}$ 
 (averaged over a unite energy window $\sim T$) 
 and of the orbitally-resolved occupation with gate voltage $V_g$ (electron
doping)  for the representative Mn sites  of the $N\!=\!46$ (upper panels) and $N\!=\!4$ (lower panels) clusters.
The red (gray) blue (dark gray) and green (light gray) curves correspond 
to representative Mn$1(1)$, Mn$1(2)$, and Mn$2$ atoms, respectively,
while the light and dark colors denote $3z^2-r^2$ and $x^2-y^2$ states.
The gate voltage $V_g$ drives an orbital selective Mott transition,
which becomes sharper upon decreasing the system size. }
 \label{fig:gateMn46}
\end{figure*}

\subsection{Site- and orbital-selective Mott transition driven by applied gate voltage} 
\label{sec:gate}

The interesting insulator-to-metal-to-insulator transition in LCMO upon size reduction 
is based on a complex interplay between quantum confinement effects and the structural distortions 
occurring upon size reduction from the bulk LCMO to $3$~nm nanoclusters. 
As the onset of the peculiar charge- and orbital-ordered state found in the bulk LCMO 
relies on the balance between Mn$^{3+}$ and Mn$^{4+}$, it is interesting 
to study the effect of electron doping on the few-atom clusters. 
A change in the number of carriers without changing the chemical composition of the system 
can be achieved by the application of an external gate voltage $V_g$. 
In this section, we investigate this issue through DMFT calculations considering the clusters
with $N\!=\!46$ and $4$ Mn atoms. We consider the limiting case 
where there is only an infinitesimally small tunneling contact with the environment, 
so that the we can account for the gate voltage by changing the DMFT chemical potential. 
We neglect the effect of doping on the DFT effective potential, as in the virtual crystal approximation. \cite{nordheimAP9}

\noindent In Fig.\ \ref{fig:gateMn46} we show the orbitally-resolved spectral weight 
at the Fermi energy $\overline{A(E_F)}$  (averaged over a unite energy window $\sim T$) 
as calculated from the Green function at imaginary time $\tau\!=\!\beta/2$, 
and the occupations of the two $e_g$ orbitals for representative sites 
of $S_{\rm{nano}}$ with $N\!=\!46$ and $N\!=\!4$ Mn sites clusters. 
The value $V_g\!=\!0$ corresponds to the results in the previous sections 
and to an average cluster occupation $\langle n \rangle\!=\!0.5$ electrons in the $e_g$ orbitals. 
Upon changing  $V_g$, we increase the number electrons in the LCMO cluster (electron doping). 
When the low-lying $e_g$ orbital (e.g., the $x^2-y^2$ in the case of Mn$1$) on a Mn site becomes half-filled, 
strong electronic correlations drive an orbital-selective metal-to-insulator transition, with the opening of a Mott gap. 
Such an orbital-selective Mott transition has been reported before for the Hubbard model, 
originating from different bandwidth (or correlations) 
for different orbitals, \cite{demediciPRB72,liebschEPL97,kogaPRL92,aritaPRB72,knechtPRB72,jakobiPRBs8087}
or due to the band degeneracy lifting, \cite{demediciPRL102} as well as for materials. \cite{anisimovEPJJB25,naupanePRL103,yuPRL110} 
The charge disproportionation between Mn$^{3+}$ and Mn$^{4+}$ also results in a strong site-selective character 
of the transition. Site-selective behavior of similar kind has also been reported recently for bulk systems 
\cite{parkPRL109,karolak1501.03294}. 
In our case, we have one insulating orbital which is integer filled ($n=1$), 
but neither the occupation of the other (metallic) orbital, nor the cluster average electron density is integer. 
One possible interpretation of this novel orbital- and site-selective Mott transition 
is associated to the role of the Hund's exchange coupling to the $t_{2g}$-spins. 
A filling of one electron in an orbital thus means that all states with spin parallel to the $t_{2g}$-spins 
are occupied, while those with opposite spin are empty. 

The orbital-selective MIT is relatively homogeneous in the cluster, 
meaning that $e_g$ Mn sites of the same kind turn insulating at a similar value of $V_g$. 
On the other hand, the critical value of $V_g$ still depends both on the kind of Mn site and the cluster size. 
Further increasing $V_g$ increases the population of the other $e_g$ orbital (e.g., the $3z^2-r^2$ in the case of Mn$1$). 
Above $V_g\!\approx\!2.6$~eV (until the chemical potential lies within the Mott gap) 
all orbitals are half-filled and Mott insulating, with $A(E_F)\!=\!0$ (see upper panels in  Fig.\  \ref{fig:gateMn46}) 
and display a divergent imaginary part of the DMFT self-energy (not shown).  
In general, we observe that, upon decreasing the cluster size, 
the orbital-selective MIT is found at a smaller value of $V_g$. 
The transition also appears to be sharper upon changing $V_g$. 
This can be understood by considering the more localized nature of the $e_g$ orbitals and 
the enhanced orbital polarization observed for smaller cluster sizes. 
This effect is important in view of possible applications. 
For an appropriate system size, half-doped LCMO manganites nanocluster 
can be driven across a MIT by applying an external gate voltage. 

\section{Conclusions}
\label{sec:conclusions}

We investigated the effects of size reduction on the charge and orbital order   
in the LCMO mixed valence manganite within the  DFT+DMFT framework. 
As was shown before, \cite{dasPRL107} the size reduction from bulk to a $3$~nm nanoclusters 
weaken the distortions the bulk crystal structure  and induces an insulator(bulk)-to-metal($3$~nm) transition 
in the high-temperature paramagnetic phase, along with a weakening of charge and orbital disproportionation. 
Here,  we extend the analysis by considering  nanoclusters of just a few-atoms.
Upon reducing the system size we observe the opposite trend: 
driven by the quantum confinement, there is a second
metal($3$~nm)-to-insulator(few atoms) transition and an enhancement of both charge and orbital disproportionation.
We also investigated the effect of electron doping on the few-atom nanoclusters by applying an external gate voltage. 
We observe an orbital-selective Mott transition at a critical value of the gate voltage which
corresponds to an integer filling of only an individual $e_g$ orbital. 
The orbital-selective nature of the transition is a direct consequence of the orbital polarization 
and the strong Hund's exchange splitting. At the same time, 
the strong charge disproportionation between Mn$^{3+}$ and Mn$^{4+}$ sites
induces also a site-selective character, with different Mn kinds turning insulating 
at different values of the gate voltage. 

Our theoretical prediction of a reentrant insulator-to-metal-to-insulator transition 
and the reported gate voltage control calls for further experiments. 
Technical applications are discernible since these two control parameters should allow us 
to fine tune LCMO nanoclusters to the verge of a Mott transition. 
In this situation smallest changes in temperature, voltage, magnetic field etc.\ 
can trigger a gigantic change in conductance.

\begin{acknowledgements}
We acknowledge financial support from the Austrian Science Fund (FWF) through I-610-N16 (AV), 
the Deutsche Forschungsgemeinschaft FOR 1346 (GS),  
and the European Research Council under the European Union's Seventh Framework Program 
FP7/ERC through grant agreement n.~306447 (AV, KH) and n.~240524 (AV). 
HD and TSD would like to acknowledge Department of Science and Technology, India for the support. 
The numerical calculations were performed on the Vienna Scientific Cluster (VSC). \\
\end{acknowledgements}



\begin{thebibliography}{99}
\bibitem{note_intpar} This relation holds in the case of degenerate orbitals, with the Hamiltonian being invariant under orbital rotations. 
However, it is considered physically sensible also when the symmetry is fulfilled only approximately, \cite{heldAP56} 
e.g., if the $e_g$ orbitals are split by the Jahn-Teller crystal field.
\bibitem{note_CPA} Note that in general the Hund's exchange ${\cal J}$ entails effective non-local (short-range) 
magnetic correlations between different Mn atoms, which are nengelected in DMFT. 
However, it has been shown\cite{rotterEPJB86} that those correlations are less important 
at high-temperature as well as for lattice structures with high-coordination, and DMFT provides reliable results. 
Also note that the average of an Ising spin $S=\pm1$ yields the same DMFT Green function 
for the paramagnetic phase as the angular average of a classical spin with $|\vec{S}|=1$.
\end{thebibliography}
\end{document}